\documentclass[aps,prl,twocolumn,showpacs,groupedaddress]{revtex4}  %
\usepackage{graphicx}  %
\usepackage{dcolumn}   %
\usepackage{bm}        %
\usepackage{amssymb}   %

\begin{document}

\newcommand{\bfs}[1]{\noindent {\large \sf \bfseries #1}}
\newcommand{\bi}{\vspace{-0.1in} \begin{itemize} \itemsep=0pt \parsep=0pt}
\newcommand{\ei}{\end{itemize}}
\newcommand{\be}{\vspace{-0.1in} \begin{enumerate} \itemsep=0pt \parsep=0pt}
\newcommand{\ee}{\end{enumerate}}

\newcommand{\bmc}{\begin{multicols}{2}}
\newcommand{\emc}{\end{multicols}}

\newcommand{\bc}{\begin{center}}
\newcommand{\ec}{\end{center}}

\newcommand{\bq}{\begin{quote}}
\newcommand{\eq}{\end{quote}}

\newcommand{\bea}{\begin{eqnarray}}
\newcommand{\eea}{\end{eqnarray}}

\newcommand{\beas}{\begin{eqnarray*}}
\newcommand{\eeas}{\end{eqnarray*}}

\newcommand{\beac}{\begin{eqnarray}}
\newcommand{\eeac}{\end{eqnarray}}

\newcommand{\beq}{\begin{equation}}
\newcommand{\eeq}{\end{equation}}

\title{\bf A simple theory of protein folding kinetics}
\author{Vijay S. Pande}
\affiliation{Departments of Chemistry, Structural Biology, and Computer Science, Stanford University, Stanford, CA }
\date{\today}

\begin{abstract}
\noindent We present a simple model of protein folding dynamics that captures key qualitative elements recently seen in all-atom simulations.  The goals of this theory are to  serve as a simple formalism for gaining deeper insight into the physical properties seen in detailed simulations as well as to serve as a model to easily compare why these simulations suggest a different kinetic mechanism than previous simple models.  Specifically, we find that non-native contacts play a key role in determining the mechanism, which can shift dramatically as the energetic strength of non-native interactions is changed.  For protein-like non-native interactions, our model finds that the native state is a kinetic hub, connecting the strength of relevant interactions  directly to the nature of  folding kinetics. 
\end{abstract}

\maketitle

{\noindent \em Introduction.---}  Protein folding has been an important problem at the crossroads of statistical mechanics, computer simulation, and biophysics.
There has been a long history of theoretical advances in the study of protein folding, and we refer the reader to reviews \cite{PandeRMP2000,DillReview,FunnelPaper,NoeMSMReview}.  Recent advances in computer simulations have enabled one to use detailed, atomistic models to simulate the complete process of folding, on relatively long (millisecond) timescales \cite{VoelzNTL92010}.  This has become possible due to the advent of Markov State Models (MSMs) (see Refs. \cite{MSM,NoeMSMReview} for recent reviews), an approach which uses detailed simulation to construct a Master equation for the statistical dynamics of a particular protein.  

By examining and comparing MSMs for different proteins (as well as by direct examination of simulations), some surprises have emerged.  Perhaps most importantly, the role of non-native contacts has now been highlighted as a key part of protein folding \cite{VoelzNTL92010,GregHub}; in hindsight, this is natural, since amino acid interactions are not particularly specific and there often is little free energetic difference between say a native-like interaction between two aromatic residues vs a non-native like interaction  \cite{VoelzNTL92010}.  This opens a new door to re-examine simple models of protein folding and to develop a new theoretical formalism to more naturally include non-native interactions.

In this work, we take a Master equation approach to dynamics, much like one does computationally with MSMs.  The key question is how to model the rate matrix.  Previous  models for protein folding kinetics \cite{Wolynes,Shakhnovich} (derived from models of spin glass dynamics \cite{REMdynamics}) have made very simple approximations for the nature of the rate matrix.  Here, we develop a new theoretical framework for the rate matrix which allows for a more detailed model of kinetics, especially the natural inclusion of non-native interactions.  The application of this model to various regimes of protein-sequences allows one to make a direct connection to recent simulations \cite{VoelzNTL92010} and, via a simple, solvable analytic model, describe the essence of {\em native hubs} (i.e. transitions between non-native states occur via the native state) recently seen in simulation \cite{GregHub}.

\medskip

{\noindent \em Model.---}  We first introduce a phenomenological 
Hamiltonian for the energy of structure $\alpha$:
\beq
{\cal H}_\alpha =  \epsilon_N \sum_{ij} C_{ij}^\alpha C_{ij}^N +
\epsilon_{NN} \sum_{ij} C_{ij}^\alpha (1-C_{ij}^N)
\eeq
where $C_{ij}^\alpha$ is the contact map of structure (microstate) $\alpha$ (i.e. either 1 if a contact between residues $i$ and $j$ is present in the structure $\alpha$ or 0 otherwise) and $C_{ij}^N$ is the contact map of the native state.  We choose the native contacts and non-native contacts to have differing energetic contributions ($\epsilon_N$ and $\epsilon_{NN}$, respectively), noting that these quantities could naturally be negative (especially $\epsilon_N$).  Note that we are including terms such as solvent entropy in the ``energy'' term above.  

This Hamiltonian reduces to ${\cal H}_\alpha = \epsilon_N q^{\alpha,N} + \epsilon_{NN} (q^{\alpha,\alpha} - q^{\alpha,N}) = \epsilon_x q^{\alpha,N} + \epsilon_{NN} q^{\alpha,\alpha}$,  where $\epsilon_x \equiv \epsilon_N - \epsilon_{NN}$ is the extra energetic preference for native over non-native contacts and $q^{\alpha,\beta} \equiv \sum_{ij} C^\alpha_{ij} C^\beta_{ij}$ is the number of contacts in common between structures $\alpha$ and $\beta$ (also note that thus $q^{\alpha,\alpha }$ is shorthand for the total number of contacts in structure $\alpha$).  This Hamiltonian is simple, yet captures the main element of interest in this work: the interplay between native and non-native interactions.

However, in order to make a connection to more detailed calculations, it is useful to note that a similar expression can be derived from more direct physical grounds, i.e. from a microscopic Hamiltonian with the explicit concept of sequence design \cite{PandeRMP2000}.  Starting from the more general, microscopic contact Hamiltonian
${\cal H}_\alpha = \sum_{ij} C^\alpha_{ij} B_{s_i,s_j}$
, where $B_{IJ}$ is a general matrix of monomer-monomer interactions (and we use capital letters to designate the space of different types of residues).
This is analogous to problems that have already been solved \cite{PandeRMP2000}, yielding (to lowest order in $1/T_d$) the effective Hamiltonian for $\alpha$ is
\beq
{\cal H}^{\rm eff}_\alpha = {\cal F}_\alpha + T S^{\rm loop}_\alpha =
\overline{B} q^{\alpha \alpha} -  {\overline{B^2_c} \over 2 T_d} q^{\alpha N}
\eeq
where $T_d$ is the design temperature (lower $T_d$ means better optimized sequences), and we see terms involving the mean ($\overline{B}$) and the variance ($\overline{B^2_c}$) of the $B_{IJ}$ matrix take the roles of of $\epsilon_{NN}$ and $\epsilon_x$, respectively, in our phenomenological Hamiltonian.  Having both representations (i.e. the phenomenological  as well as the microscopic Hamiltonian) allows us also to make a natural connection to previous work \cite{PandeRMP2000}.  We will return to the sequence-based Hamiltonian results at the end, in order to make a more direct connection to protein biophysics.

\medskip

{\noindent \em Kinetics formalism.---}  To build a kinetic model, we consider the master equation:
\beq
{dp_{\alpha} \over dt} \ = \ \sum_{\beta }  \  k_{\alpha \beta} \ p_\beta
\label{MasterEq}
\eeq
which means that we must consider the nature of the rate matrix $k_{\alpha \beta }$.  
To more easily see the impact of these elemental rates on the overall dynamics, we propose a simple model for the rate matrix:  a block-diagonal matrix with a block diagonal form of $n$ blocks each of size $m$ rows but now with one additional row for the folded (native) state.  Specifically, with elements of the form
\beq
k_{\alpha \beta} = \left\{ 
\begin{array}{cl}
k_1 &  {\rm within \ a \ nonnative \ block } \\
k_0 &  {\rm between \ nonnative \ blocks} \\
k_{0N} &  {\rm from \ non\!-\!native  \ to \ native} \\
k_{N0} &  {\rm from \ native  \ to \ non\!-\!native} \\
- \sum_{\beta \neq \alpha} k_{\alpha \beta} &  {\rm on \ diagonal} 
\end{array}
 \right.
\label{KabformEq}
\eeq
Here, the $n$ blocks are meant to represent $n$ mestastable states, each consisting of  $m$ highly related conformations.   
Note that this matrix obeys detailed balance, although we have made the simplifying approximation that the free energy of states within a block are similar and the free energy between non-native blocks is also similar. 
 
This rate matrix has a well defined set of degenerate eigenvalues: 
a non-degenerate eigenvalue of $0$ (the equilibrium eigenvalue), a $(n-1)$-fold degenerate eigenvalue of $\kappa_0 \equiv n m k_0 + k_{N0}$ (for transitions between non-native blocks), a $[n (m -1) ]$-fold degenerate eigenvalue of $m (n - 1) k_0 + m k_1 + k_{N0})$  (for transitions within a block), and a non-degerate eigenvalue of $\kappa_N \equiv n m k_{0N} + k_{N0}$ (for transitions to the native state).  We note in passing that the degeneracy in the eigenvalues seen here is naturally broken by some small variations in rates between states (i.e. the rates between all non-native states would not be all exactly $k_0$).  Also, variations in the value of $m$ between blocks do not change the results discussed below.

In the large $n$ limit, the ratio of rates of transformations between non-native blocks vs those from non-native to native will be  $\kappa_0/\kappa_N = (n m k_0 + k_{N0})/(n m k_{0N} + k_{N0}) \approx k_0/k_{0N}$.  Thus, our primary goal will be to compare these elemental rates.  
To do so, we take a Kramer's approximation for dynamics between $\alpha$ and $\beta$, i.e.
\beq
k_{\alpha \beta } = \tilde{k} \exp[-(F_{\alpha \beta }^\ddag - F_\alpha)/T]
\label{KramersEq}
\eeq
where $\tilde{k}$ is the microscopic rate of interconversion,
$F_\alpha$ is the free energy of $\alpha $, and $F_{\alpha \beta }^\ddag$  is the free energy of the transition state (denoted by $\ddag$) between  $\alpha$ and $\beta$ (note that this is not the global transition state, but just the transition state between structures $\alpha$ and $\beta$).  Since the eigenvalues and eigenvectors of the $k_{\alpha \beta}$ matrix define the relevant timescales and dynamics, respectively, the next step is to flesh out this matrix in more detail.

There have been previous approximations to model $k_{\alpha \beta }$, notably setting  $k_{\alpha \beta } = \tilde{k} \exp(-E_\beta/T)$ \cite{REMdynamics} or $k_{\alpha \beta } = \tilde{k} \exp(E_\alpha/T)$ \cite{StretchedREMdynamics}, which yielding solvable models within the Random Energy Model (REM) leading to powerlaw \cite{REMdynamics} and stretched exponential \cite{StretchedREMdynamics} relaxation, respectively.  Also of note is an extension to GREM \cite{Wolynes}.  However, there are two key limitations to these methods.  First, they only directly apply  to a theory for kinetics of random sequences.
Second, as we argue below, by considering the structure of the transition state for transitions explictly, we can  improve upon the previous models.

We propose that the transition state between structures $\alpha$ and $\beta$ can be approximated in terms of the contacts in common between these structures:
\beq
C^\ddag_{ij} = C^\alpha_{ij} C^\beta_{ij}
\label{TSCMEq}
\eeq
(here, we take advantage of the fact that contacts are either valued at 0 or 1, so multiplication works like a binary AND operator).  Physically, this models the transition between  $\alpha$ and $\beta$ as breaking the contacts in $\alpha$ not present in $\beta$ and then forming the contacts present in $\beta$ that were not originally present in $\alpha$. 
We note in passing that this approach is potentially broadly applicable to a range of problems, whose state information can be encapsulated into a binary vector analogous to $C^\alpha_{ij}$.

Eq (\ref{TSCMEq}) is an advance over previous work in two ways.  First, we consider directly the microscopic transitions between states, i.e. not considering these transitions in terms of all going through a single barrier, but many different pair-wise barriers.  Second, we look directly to {\em structural} properties of the state to determine the nature of the transition state structure.  However, we stress that eq (\ref{TSCMEq}) is most appropriate for transitions between collapsed (or mostly collapsed) states.

To calculate the free energy as a function of a given state, we must include the energy (from the Hamiltonian above) as well as a model for the polymeric entropy.  We follow the model described in \cite{PandeKinetics1997} and say that for a chain of $N$ persistence lengths, the number of contacts present ($q^{\alpha,\alpha }$) in the structure $\alpha$ lead to $q^{\alpha,\alpha }$ loops, each of length $\ell \sim N/q^{\alpha,\alpha }$; these loops each contribute an entropy of $\Delta S_\alpha^{\rm loop}/k_B = -\sigma  + (3/2) \ln q^{\alpha,\alpha }$, where $\sigma \equiv s - (3/2) \ln N$ and $s$ is a positive quantity related to the flexibility of the chain and $k_B$ is Boltzmann's constant; note that the value of $\Delta S^{\rm loop}_\alpha$ (eg as seen in lattice model calculations \cite{PandeRMP2000}) is dominated by the $\sigma$ term.  

This leads to the transition state energy
\bea
E_{\alpha \beta }^\ddag & = & \epsilon_N \sum_{ij} C^\alpha_{ij} C^\beta_{ij} C_{ij}^N +
\epsilon_{NN} \sum_{ij} C^\alpha_{ij} C^\beta_{ij} (1-C_{ij}^N)
\nonumber \\
& = & \epsilon_x q^{\alpha,\beta,N} + \epsilon_{NN} q^{\alpha,\beta} 
\eea
where $q^{\alpha,\beta,N} \equiv \sum_{ij} C^\alpha_{ij} C^\beta_{ij} C_{ij}^N$ is the three-conformation overlap between $\alpha$, $\beta$, and the native state.
Combining terms above (and including the entropy), we get
\bea
F_{\alpha \beta }^\ddag - F_\alpha & = &
\epsilon_x ( q^{\alpha,\beta,N} - q^{\alpha,N})  \ + 
f_{NN} (q^{\alpha,\beta} - q^{\alpha,\alpha} ) + \nonumber \\
&& - (3/2) k_B T ( q^{\alpha,\beta } \ln q^{\alpha,\beta }  -  q^{\alpha,\alpha } \ln q^{\alpha,\alpha} )
\label{FtsEq}
\eea
where we have defined the effective free energy of contact formation $f_{NN} = \epsilon_{NN} + k_B T \sigma$.
With this barrier height now directly connected to properties of the Hamiltonian, we are now ready to examine specific models for folding.

\medskip

{\noindent \em Exploring the model.---}  
To examine the model, we consider some limiting cases below.  
First, we wish to consider a model for protein-like sequences, i.e. those which resulted from evolution (or alternatively protein sequence design).  In order to obtain reasonable estimates for the key transition rates in our model, i.e. $k_0 = k_{\alpha \beta}$ vs $k_{0N} = k_{\alpha N}$ (where $\alpha$ and $\beta$ are representative non-folded structures and $N$ is the native state), we look to equations (\ref{MasterEq}), (\ref{KramersEq}), and (\ref{FtsEq}).  First, we consider the case $\epsilon_N < \epsilon_{NN} < 0$, i.e. both native and non-native contacts are energetically preferred, but native more so.  The intrastate conversion rate will be the fastest rate, since the states are very similar (i.e. $q^{\alpha,\beta} \approx q^{\alpha,\alpha}$ and $q^{\alpha,\beta,N} \approx q^{\alpha,N}$), which leads to a very low barrier height from eq (\ref{FtsEq}).  In order to compare $k_{0N}$ to $k_0$, consider that both $\epsilon_N$ and $\epsilon_{NN}$ are negative, but the drive to form native contacts is stronger; thus, the barrier height determined in eq (\ref{FtsEq}) will be lower for transitions to $N$.

More specifically, it is instructive to compare the barrier transitions from  $\alpha$ to some other non-folded structure $\beta$ vs folding from $\alpha$ to $N$.  The difference in barrier heights $\Delta \Delta F^\ddag \equiv  (F_{\alpha \beta }^\ddag - F_\alpha) - (F_{\alpha N }^\ddag - F_\alpha) = \Delta \Delta E^\ddag - T \Delta \Delta S^\ddag = -k_B T \ln (k_{\alpha \beta}/k_{\alpha N})$ is given by the combination of an energetic
\bea
\Delta \Delta E^\ddag = 
-\epsilon_x ( q^{\alpha,N} - q^{\alpha,\beta,N})  - 
\epsilon_{NN} (q^{\alpha,N} - q^{\alpha,\beta} ) 
\label{DDFEq}
\eea
and entropic $\Delta \Delta S^\ddag  = (S_{\alpha \beta}^\ddag - S_\alpha) - (S_{\alpha N}^\ddag - S_\alpha)$
\bea
- T \Delta \Delta S^\ddag 
& = & 
- k_B  T \sigma (q^{\alpha,N } - q^{\alpha,\beta })  + \\
&&   (3/2) k_B T ( q^{\alpha,N } \ln q^{\alpha,N } - q^{\alpha,\beta } \ln q^{\alpha,\beta })
\nonumber
\eea 
contributions.

Consider the regime where both $\epsilon_x$ and $f_{NN}$ are negative quantities.   Note that $f_{NN}$ represents the effective free energetic drive to form contacts in general and is negative when there is a sufficient general attraction that beats out the loss of entropy of forming contacts.
Since three-body overlaps are much more rare than two-body overlaps in the unfolded state, then $q^{\alpha,N} > q^{\alpha,\beta,N}$.  Finally, due to the nature of our Hamiltonian, it would be rare for two structures at random to have more contacts in common with each other than with the native state; thus, we can make the approximation that  at least $q^{\alpha,N} \approx q^{\alpha,\beta}$, or more likely $q^{\alpha,N} \ge q^{\alpha,\beta}$.  Thus, putting this all together, we find that, for this regime, $\Delta \Delta F^\ddag$ is positive, which yields $k_{0N} > k_0$.  As the strength of non-native interactions gets more attractive (i.e. $\epsilon_{NN}$ more negative) and the difference between native and non-native strength grows (i.e. $\epsilon_N$ and $\epsilon_{NN}$ negative and $\epsilon_x<0$), the more that $k_{0N}$ is greater than $k_0$, emphasizing this effect.  

Moreover, since the different collapsed globules have roughly the same number of total contacts, the rates for unfolded state globule folding to the native state are largely uniform (justifying our model as constructed in Eq~(\ref{KabformEq})), yielding single exponential kinetics even though there are many parallel pathways.

Due to the exponential nature of the relationship between rates and free energies, this leads to the relationship $k_{0N} > k_0$.  What are the implications of this?  In this limit, we would find that folding to the native state is fast, compared to dynamics from one non-folded state to another.  This makes the folded state a {\em kinetic hub}, i.e. transitions between states are typically mediated through the native state.  Moreover, generalizations of this model with a higher level hierarchy show (as could be seen from numerical solutions of this model) that the native state is a kinetic hub, i.e. transitions between non-native states usually go through the native state.

Finally, we use the previous model to examine another regime, where native contacts dominate, i.e. $\epsilon_N$ is negative but $f_{NN}$ is positive.  In this case, native contacts are strongly preferred and non-native contacts are discouraged (eg with excluded volume repulsion and no attraction due to interactions, as is common in computer simulations of G\=o models \cite{FunnelPaper}).  
Specifically, as we see directly from eq (\ref{DDFEq}), when $f_{NN}$ is sufficiently positive, we obtain a negative value for $\Delta \Delta F^\ddag$.  Thus, this would lead to the model parameters of the form $k_0 \gg k_{0N}$.

Physically, interconversion between non-native states is fast, since they are not separated by barriers (their transition state energies have no contributions from the free energy of breaking non-native contacts).  This regime leads to a very different picture (akin to the ``smooth energy landscape'' picture previously described \cite{FunnelPaper}), in which the unfolded state interconverts quickly in general, waiting for the rare chance to fold.

\medskip

\begin{figure}[bt]
\begin{center}
\includegraphics[width=3.25in]{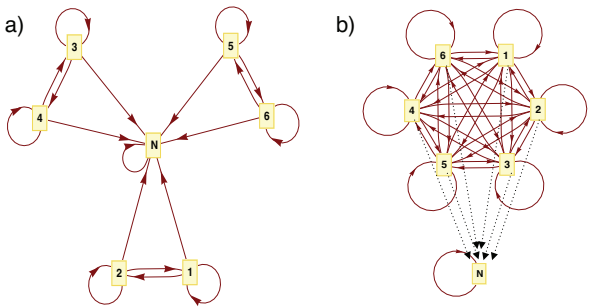}
\label{Graphs}
\end{center}
\caption{Two different kinetic regimes result from our theory, as demonstrated in a simple numerical example with $n=3$ blocks and $m=2$ states per block, plus the native state (7 states total); see eq. (\ref{KabformEq}) for details of the rate matrix structure.  The theory presented here is used to calculate the mean first passage time (MFPT) between states; edges are shown as solid lines if the MFPT is fast ($<30/\tilde{k}$), with the MFPT of an edge listed to one signficant digit.  In both examples, we set $k_{N0} = k_{0N} \exp(-8/0.6)$ and $k_1=1 \tilde{k}$.    a) For estimates based on the MJ matrix regime ($k_{0N}=0.05\tilde{k}$,  $k_0=0.001\tilde{k}$), we see that the native state (N) is a kinetic hub.  b) For a G\=o model regime ($k_{0N}=0.005\tilde{k}$, $k_0=0.5\tilde{k}$) we see that there is fast interconversion between unfolded states (1-6), with slow interconversion to the native state (shown by dotted lines).}
\end{figure}

{\noindent \em Discussion and Conclusions.---}  While previous theoretical approaches \cite{Wolynes, Shakhnovich} have made seminal contributions to the theoretical framework for understanding folding, these models did not model the transition state structurally, which has  particularly important implications for the impact of non-native interactions.  
As we have seen above, the inclusion of non-native interactions critically changes the qualitative behavior of the model; indeed, this regime has been shown to be particularly relevant in recent all-atom protein folding simulation.  

Moreover, we can derive  estimates for our parameters from previous studies of proteins.  For example, the Miyazawa and Jernigan \cite{MJmatrix} matrix's mean ($\overline{B} \approx -3.2 k_B T$) and a standard deviation  ($(\overline{B^2_c})^{1/2} \approx 1.5 k_B T$), respectively.  These results, consistent with other such estimates (such as amino acid solvation free energies \cite{MJmatrix}), combined with estimates that $k T_d \approx 0.8 k_B T$ and $k_B T s \approx 1.5 k_B T$ \cite{PandeRMP2000} indicate that $f_{NN}<0$; our theory therefore predicts that proteins would fold with the native state as a kinetic hub (i.e. fast folding to the native state, compared to equilibration between unfolded states) \cite{GregHub}, depicted in Fig.~1.   These averages are formally weighted by the amino acid composition \cite{PandeRMP2000} of the protein sequence and a uniform composition 
is used above.

We also note in passing that while an overall collapsed model is handled by our theory, the limiting case of minimal (or repulsive) non-native interactions is not, since that regime would not lead to collapsed configurations and thus eq (\ref{TSCMEq}) would not be valid; however, this regime is already well understood: the preponderance of contacts are native in this case, and thus folding proceeds by the formation of these contacts, as seen  previously \cite{PandeRMP2000}.

With this new formalism, we are able to recover and potentially explain the behavior seen in all-atom simulations \cite{VoelzNTL92010, GregHub}.  Specifically, we get the primary result that the dynamics of interconversion from one non-folded state to another can be very slow.    This also leads to a secondary result that native state is a kinetic hub when there is some non-native attraction.   This suggests that simple, previous choices for a single dimensional reaction coordinate  (such as using the number of native contacts) can lead to a misconception in terms of the fundamental dynamics of proteins, since these approaches assume that the unfolded state is rapidly interconverting.  
This is correct for some G\=o models of protein folding, but not for models which include non-native attraction.  

Finally, we stress that the property of the unfolded state predicted from this theory does not apply to the chemically {\em denatured} state, in which most experiments probing the ``unfolded state'' of proteins have been performed; our theory predicts that experiments directly examining the true unfolded state will see a much slower relaxation time compared to the denatured state.

To conclude, one of the motivations of this work was to develop a model which was simple enough that it could be solved analytically, but with the key essence of protein folding seen in detailed simulations.  The  qualitative change which derives from the simple addition of the role of non-native contacts shows how this model can easily be used to probe folding dynamics.  By combining detailed simulations with  analytic approaches,  insight in a single system studied by simulation could be extended to a broad range of proteins and protein folding phenomena.

\noindent {\em Acknowledgements.---}  We thank S. Bacallado, K. Beauchamp, G. Bowman, J. Chodera, J. England, A. Grosberg, P. Kasson, M. Levitt, L. Maibaum, and V. Voelz for discussions and  NSF (EF-0623664) and NIH (R01-GM062868) for funding.

\end{document}